\documentclass[11pt,a4paper,leqno]{article}
\title{On the  variational problem for the upper bounds
of solute transport in  double-diffusive convection}
\author{Zlatinka Dimitrova$^1$, Nikolay Vitanov$^2$ \footnote{corresponding author}}
\date{}
\begin{document}
\maketitle
\begin{abstract}
The formulation of the variational problems for the solute transport
in a fluid layer in presence of double-diffusive thermal convection is discussed.
It is shown that the variational functional obtained by Strauss can be generalized
and the general functional leads to accurate upper bounds on the solute
transport for the case of small and intermediate values of the Rayleigh number.
The general functional however is a non-homogeneous one  but for asymptotically
large Rayleigh numbers it converges to the Strauss approximation. Thus for
small and intermediate values of the Rayleigh numbers one should use the general
functional and for vary large values of the Rayleigh numbers one can use the
functional of Strauss. 
\end{abstract}
Double-diffusive comvection is of considerable interest for oceanology
as it is connected to the convective motion of water containing some
amount of salt. In this case of convection in addition to the temperature
gradient there is also a gradient of the salt concentration and this
complicates the model equations. Below we discuss the variational problem
for the upper bounds on the solute transport in a double diffusiove convection.
Our discussion will be based on the optimum theory of turbulence which has been
developed in the pioneering works of Howard, Busse, Doering and Constantin $^{1-4}$. 
We shall not discuss the Doering-Constantin approach which easily leads to
upper bounds if appropriate background fileds are chosen$^{5}$. Instead of this
our attention will be concentrated on a discussion of the work of Strauss$^{6}$
which is based on the Howard-Busse method which has many application to
systems with thermal convection$^{7-11}$. In order to avoid the complication of
the analysis we shall not discuss the case of presence of rotation$^{12-16}$ and
instead of this we shall follow the giudelines from $^{17,18}$.
\par
Below we shall model the double-diffusive convection in a horizontal fluid layer
by means of the equations of Boussinesq approximation$^{6,19}$
\begin{equation}\label{e1}
\frac{\partial \vec{u}}{\partial t} + \vec{u} \cdot \nabla \vec{u} =
- \frac{1}{\rho_0} \nabla p - \beta g S \vec{e}_z + \gamma g T \vec{e}_z
+ \nu \nabla^2 \vec{u} 
\end{equation}

\begin{equation}\label{e2}
\frac{\partial {\cal S}}{\partial t} + \vec{u} \cdot \nabla {\cal S} = \kappa_{S} \nabla^2 {\cal S} 
\end{equation}

\begin{equation}\label{e3}
\frac{\partial {\cal T}}{\partial t} + \vec{u} \cdot \nabla {\cal T} = \kappa_{T} \nabla^{2} {\cal T}
\end{equation}

\begin{equation}\label{e4}
\nabla \cdot \vec{u} =0
\end{equation}

where ${\cal S}$ is the concentration field which is divides as a horizontally averaged part ${\overline {\cal S}}$
\footnote{In this paper we shall use two kinds of averages:
\begin{enumerate}
\item
Average over horizontal plane (horizontal average)
$$
\overline{Q}(z,t) = \lim_{L \to \infty} \frac{1}{4 L^2} \int_{-\infty}^{\infty} \int_{-\infty}^{\infty} dx dy \ \ Q (x,y,z,t)
$$
\item
average over the volume of the layer (volume average)
$$
\langle{Q} \rangle(t) = \lim_{L \to \infty} \frac{1}{4 L^2} \int_{-\infty}^{\infty} \int_{-\infty}^{\infty} \int_{0}^{1}  dx dy dz \ \ Q (x,y,z,t)
$$
\end{enumerate}} \footnote{
In additition we shall assume that we investigate the process of convection under quasistationary condition
which is defined by the requirement that the horizontal averages of all quantities are time independent and
the horizontal averages of all fluctuation parts of the quantities are $0$. Because of this in the text below
the horizontal and volume averages of the studied quantities depend only on $z$ and the horizontal averages
of the fluctuating parts of the studied quantities are set to $0$.
}
and fluctuating part $S$

\begin{equation}\label{e5}
{\cal S} (x,y,z,t) = \overline{{\cal S}} (z) + S (x,y,z,t)
\end{equation}
The same division is performed for the temperature field ${\cal T}$ too
\begin{equation}\label{e6}
{\cal T} (x,y,z,t) = \overline{{\cal T}} (z) + T (x,y,z,t)
\end{equation}
In the above equations $\vec{u}=(u,v,w)$ is the fluid velocity (which is a fluctuating quantity as we assume
that there is no mean flow: ${\overline{\cal U}} =0$ \footnote{The basis of this assumption is in the
nature of the convective motion in a fluid layer. When the heating of the layer is not very intensive
the heat is transported through the layer by means of thermal conduction. Thermal conduction means that
there is some temperature distribution in the fluid layer ( i.e., $\overline{{\cal T}} \ne 0$ and the temperature fluctuations are
zero) but there is no motion of fluid (i.e. the mean fluid velocity $\overline{{\cal U}} = 0$) and the fluctuations of the
fluid velocity are $0$ too. The arising of the convective motion of the fluid complicates the heat transfer.
The heat is transferred not only by thermal conduction but also by thermal convection. This means that non-zero
fluctuations arise and in addition the amplitude of these fluctuations can be large: they are nonlinear
fluctuations which in general can not be treated by linear model equations. Because of this the obtained
below model Euler-Lagrange integral-differential equations will be nonlinear.}). $\rho_0$ is the average fluid
density; $p$ is the deviation of the pressure from hydrostatic pressure field which would exist if
${\cal S} = \overline{\cal{S}}$ and ${\cal T} = \overline{\cal{T}}$; $\gamma$ is the coefficient of thermal
expansion; $\beta$ is the density change due to the unit change of solute concentration; $g$ is the
acceleration of the gravity; $\nu$ is the kinematic viscosity; $\kappa_{S}$ is the solute diffusivity;
$\kappa_{T}$ is the thermal diffusivity and $\vec{e}_z$ is the unit vector in the direction opposite
to the direction of gravity.
\par
Below we shall obtain several integral-differential consequences of the Boussinesq equations.
In the proces of obtaining of these relationships we shall use actively the condition of
incompressibility of the fluid (the continuity equation (\ref{e4})) and we the assumption that we
study the system long after the last external influence of its dynamics, i.e., when the condition
of quasistationarity is fulfilled that leads to independence of the horizontal averages of the
flow quantitities on the times and in addition leads to the condition that the horizontal averages
of any fluctuation quantities (and of combinations and derivatives of such quantities) are equal
to $0$.
\par
Taking into account the above mentioned conditions after averaging (\ref{e2}) with respect
to horizontal plane we obtain the relationship
\begin{equation}\label{e7}
\frac{d}{dz} \overline{w {\cal S}} = \kappa_S \frac{d^2 {\cal S}}{dz^2}
\end{equation}
The horizontal average of (\ref{e3}) leads to
\begin{equation}\label{e8}
\frac{d}{dz} \overline{w {\cal T}} = \kappa_T \frac{d^2 {\cal T}}{dz^2}
\end{equation}
The integration of (\ref{e7}) and (\ref{e8}) with respect to $z$ leads to the
relationships
\begin{equation}\label{e9}
\overline{w S} - \kappa_S \frac{d {\cal S}}{dz} = {\rm const}
\end{equation}
\begin{equation}\label{e10}
\overline{w T} - \kappa_T \frac{d {\cal T}}{dz} = {\rm const}
\end{equation}
The volume average of (\ref{e9}) and (\ref{e10}) leads to
\begin{equation}\label{e11}
\langle w S \rangle - \kappa_S \Delta S = {\rm const} \cdot d
\end{equation}
\begin{equation}\label{e12}
\langle w T \rangle - \kappa_T \Delta T = {\rm const} \cdot d
\end{equation}
where $\Delta S$ and $\Delta T$ are the differences in the salt concentration
and temperature difference between the top and bottom borders of the fluid layer.
As the integration constants are the same in (\ref{e9}) and (\ref{e11}) as
well as in (\ref{e10}) and (\ref{e12}) we can eliminate the constants and in such
a way we arrive at the two relationships 
\begin{equation}\label{e13}
\overline{w S} - \kappa_S \frac{d {\cal S}}{d z} = \frac{\langle w S \rangle}{d} - \kappa_S \frac{\Delta S}{d}
\end{equation} 
\begin{equation}\label{e14}
\overline{w T} - \kappa_T \frac{d {\cal T}}{d z} = \frac{\langle w T \rangle}{d} - \kappa_T \frac{\Delta T}{d}
\end{equation} 
Below we shall perform non-dimensionalization of the quantities on the basis of the
following units: $\Delta T$ will be the unit for temperature; $\Delta S$ will be the unit
for solute concentration; the layer thickness $d$ will be the unit for length; $d^2/\kappa_S$ will
be the unit for time and $\kappa_S/d$ will be the unit for velocity.
\par
As next step we shall ontain two relationships known also as power integrals.
Let us multiply (\ref{e2}) by $S$ and average over the fluid layer. The non-
dimensional result is
\begin{equation}\label{e15}
\langle wS \rangle^2 - \langle \overline{w S}^2 \rangle = \langle w S \rangle + \langle \mid \nabla
S \mid^2 \rangle
\end{equation}
The second power integral is obtained after multiplication of (\ref{e1}) by $\vec{u}$ averaging
the obtained result over the fluid layer and non-dimansionalization of what is obtained.
Thus we arrive at the relationship
\begin{equation}\label{e16}
- R \langle w S \rangle + \frac{\kappa_S}{\kappa_T} Ra \langle w T \rangle - 
\langle \mid \nabla \vec{u} \mid^2 \rangle =0
\end{equation}
where $R = \frac{\beta g \Delta S d^3}{\nu \kappa_S}$ is the solute Rayleigh number
and $Ra=\frac{\gamma g \Delta T d^3}{\nu \kappa_S}$ is the thermal Rayleigh number.
\par
The goal of the methodology is  the obtaining of upper bounds on the solute transport
\begin{equation}\label{e17}
Nu_S = 1 + \langle - w S \rangle
\end{equation}
We shall write $Nu_S$ in the form $Nu_S = 1 + \frac{1}{{\cal F}}$, i.e., when $Nu_S$
has a maxumum ${\cal F}$ will have a minimum. In order to come to such a form of the
relationship for the solute transport we rescale $w$, $S$, and $\theta'=\frac{\kappa_S}{\kappa_T} T$ 
as follows
\begin{equation}\label{e18}
w= A \hat{w}; \hskip.5cm S = B \hat{S}; \hskip.5cm \theta' = A \hat{\theta} 
\end{equation}
where $A$ and $B$ can be determined by substitution of (\ref{e18}) in the two power
integrals above. The result of this is
\begin{equation}\label{e19}
B = A \frac{Ra \langle \hat{w} \hat{\theta} \rangle - \langle \mid \nabla \vec{\hat{u}} \mid^2 \rangle}{R
\langle \hat{w} \hat{S} \rangle} = \alpha A
\end{equation}
\begin{equation}\label{e20}
A = \left [ \frac{\langle \hat{w} \hat{S} \rangle + \alpha \langle \mid \nabla \hat{S}
\mid^2 \rangle }{\alpha (\langle \hat{w} \hat{S} \rangle^2 - \langle \overline{\hat{w} \hat{S}}^2 
\rangle)} \right ]^{1/2}
\end{equation}
Now after some calculations we obtain the following relationship for ${\cal F}$ (we have
set $S = \tilde{S}/R$ and in the final result below we  omit the tilde and the hat signs)
\begin{equation}\label{e21}
{\cal F} = \frac{\lambda \langle \mid \nabla  S \mid^2 \rangle [\langle \mid \nabla \vec{u} \mid^2 \rangle -
Ra \langle w \theta \rangle] + (\langle \overline{w S}\rangle - \langle w S \rangle)^2}{\langle w S \rangle^2}
\end{equation}
On the basis of (\ref{e21}) we shall formulate the variational problem below.
But before this we shall obtain the particular case of (\ref{e21}) discussed by Strauss$^{6}$.
In this particular case there is a relationship between $w$ and $\theta$ and because of
this the variational problem becomes simpler and the number of the corresponding
Euler - Lagrange equations decreases. In order tho obtain the relationship between the
two above-mentioned fields we start from Eq. (\ref{e3})  perform a non-dimensionalization and discuss
the quasi-stationary approximation. What is obtained is
\begin{equation}\label{e22}
\tau w \frac{d \overline{{\cal T}}}{d z} + \tau \vec{u} \cdot \nabla T = \frac{d^2 \overline{{\cal T}}}{dz^2}+ \nabla^2  T
\end{equation}
where $\tau = \kappa_S/\kappa_T$. Here we shall use an approximate relationship for $\frac{d \overline{{\cal T}}}{d z}$
which will lead to the approximation used by Strauss. We start from (\ref{e14}) perform a non-dimensionalization
and obtain
\begin{equation}\label{e23}
\frac{d \overline{{\cal T}}}{dz} = 1 - \frac{\kappa_S}{\kappa_T} [\langle w T \rangle - \overline{\langle w T} \rangle]
\end{equation}
Now the assumption of Strauss is that $\tau = \kappa_S/\kappa_T <<1$ and that the expression in $[...]$ of
Eq. (\ref{e23}) is not larger than $1$ for any $z$. Then
\begin{equation}\label{e24}
\frac{d \overline{{\cal T}}}{dz} \approx 1
\end{equation}
and after the approximation $\tau \vec{u} \cdot \nabla T \approx 0$ the Eq. (\ref{e22}) can be reduced to
\begin{equation}\label{e25}
\tau w = \nabla^2 T
\end{equation}
and because of the fact that $\theta' = T/\tau$ this lead to
\begin{equation}\label{e26}
w = \nabla^2 \theta'
\end{equation}
Let $\theta = Ra \theta'$ and we consider the particular case of $1-\alpha$ solution of the
variational problem
\begin{equation}\label{e27}
w= w(z) f(x,y); \theta = \theta(z) f(x,y); \hskip.5cm \nabla_1^2 f = -\alpha^2 f 
\end{equation}
Then from (\ref{e26}) we obtain
\begin{equation}\label{e28}
\theta(z) = - \frac{w(z) Ra}{\alpha^2}
\end{equation}
Then the relationship for ${\cal F}$ in the approximation of Strauss is
\begin{equation}\label{e29}
{\cal F} = \frac{\lambda \langle \mid \nabla  S \mid^2 \rangle [\langle \mid \nabla \vec{u} \mid^2 \rangle +
\frac{Ra}{\alpha} \langle w^2 \rangle] + (\langle \overline{w S}\rangle - \langle w S \rangle)^2}{\langle w S \rangle^2}
\end{equation}
Note that the functional of Strauss (\ref{e29}) is homogeneous one which will lead to
homogeneous Euler-Lagrange equations and will simplify their numerical and analytical
asymptotic solutions.
\par
Let us now discuss a more general approximation and its consequences for the variational
problem. let us again assume $\tau \vec{u} \cdot \nabla T \approx 0$ and let us make Eq.(\ref{e8})
dimensionless. The result is
\begin{equation}\label{e30}
\tau \frac{d}{dz} \overline{wT} = \frac{d^2 {\cal{T}}}{dz^2}
\end{equation}
Thus after a substitution of Eqs. (\ref{e23}) and (\ref{e30}) in Eq. (\ref{e22})
we obtain the following relationship
\begin{equation}\label{e31}
\tau w \{1 - \tau [\langle w T \rangle -  \overline{wT}] \} = \tau \frac{d}{dz} \overline{wT} + \nabla^2 T
\end{equation}
Now we see that one can obtain the approximation of Strauss if one neglects the terms of
order $O(\tau^2)$ in the left-hand side of Eq. (\ref{e31}) and in addition one has to make
so-called internal layer approximation in the right-hand side of Eq.(\ref{e31}). The internal layer approximation
means that one has to assume that $\frac{d}{dz} \overline{wT} \approx 0$ in the fluid layer. Strictly speaking
the internal layer approximation is valid only in the internal sub-layer of the fluid layer. Up to some extent
it is valid in the two intermediate layers of the fluid layer \footnote{The intermediate layers of the fluid
layer are layers where the velocity field, the temperature field, and the concentration field make a transition
from their almost constant values in the internal layer of fluid to the sharply depending on the coordinate
values in the boundary layers of the fluid.}. The internal layer approximation is not valid in the boundary
layers of the fluid where all the fields (velocity, temperature and concentration) have to adjust their values
to the boundary conditions on the borders of the fluid layer. When the Rayleigh numbers connected to the
discussed problem have small and intermediate values then the thickness of the boundary layers of the
fluid layer is large and because of this the approximation of Strauss is very crude. But when the
values of the Rayleigh numbers become larger and larger then the boundary layers of the different fields
become thinner and thinner and the approximation of Strauss can became reasonable one.
\par
On the basis of all this we can formulate the variational problem for the upper bounds on the
double-diffusive convection as follows\\
\begin{center}
{\bf Case of arbitrary values of the Rayleigh numbers (Variational problem 1)}
\end{center}
{\sl
Given $Ra$ and $\lambda > 0$, find the minimum $M(\lambda)$ of the functional:
\begin{eqnarray*}
{\cal F} = \frac{\lambda \langle \mid \nabla  S \mid^2 \rangle [\langle \mid \nabla \vec{u} \mid^2 \rangle -
Ra \langle w \theta \rangle] + (\langle \overline{w S}\rangle - \langle w S \rangle)^2}{\langle w S \rangle^2} +
\nonumber \\
\mu \left \{  w - \frac{1}{Ra} \nabla^2 \theta - \frac{\tau}{Ra} \overline{w \theta} + 
\frac{\tau^2}{Ra} w [\langle w \theta \rangle - \overline{w \theta}] \right \}
\end{eqnarray*}
within the class of fields satisfying the boundary conditions and the continuity equation
$\nabla \cdot \vec{u} =0$.
}\\
Above $\mu$ is a Lagrange multiplier by means of which one takes into an account Eq.(\ref{e31}).
\par
For asymptotic large values of the Rayleigh numbers we can use the approximation of Strauss:
\begin{center}
{\bf Case of asymptotic large values of the Rayleigh numbers (Variational problem 2 - approximation of Strauss)}
\end{center}
{\sl Given $Ra$ and $\lambda > 0$, find the minimum $M(\lambda)$ of the functional:
$$
{\cal F} = \frac{\lambda \langle \mid \nabla  S \mid^2 \rangle [\langle \mid \nabla \vec{u} \mid^2 \rangle +
\frac{Ra}{\alpha} \langle w^2 \rangle] + (\langle \overline{w S}\rangle - \langle w S \rangle)^2}{\langle w S \rangle^2}
$$
within the class of fields satisfying the boundary conditions and the continuity equation
$\nabla \cdot \vec{u} =0$.}
\par
As concluding remarks we make several notes with respect to the two variational problems.
\begin{itemize}
\item
The more general variational problem 1 is a non-homogeneous one. Because of this the
corresponding Euler-Lagrange equations will be non-homogeneous too which means that
the task for the obtaining  upper bounds on the solute transport will be very difficult.
However the obtained bounds will be very accurate especially for the case of small
values of the Rayleigh numbers where the Strauss approximation is relatively inaccurate
because of the large thickness of the boundary layers of the fields of the velocity, temperature
and concentration.
\item 
For large values of the Rayleigh numbers the upper bounds connected to the more general
variational problem 1 will come close to the upper bounds of the less general
variational problem 2 of Strauss and then the more simple and homogeneous
Euler - Lagrange equation connected to the variational problem of Strauss
can be used for calculation of the upper bounds.
\item
The variational problem of Strauss is derived only for the simplest possible form of the
studied fields: fields with one characteristic wavenumber. The more general variational
problem 1 is suitable for investigation of bounds for optimum fields that have arbitrary
number of characteristic wave-numbers (so called multi-wave-number solutions of the Euler - Lagrange
equations.
\end{itemize}

\begin{flushleft}
{\bf Acknowledgment}
\end{flushleft}
This research was partially supported by the grant DO 02-338/22.12.2008 with the
Fond for Scientific Ressearches of Republic of Bulgaria.
\begin{center}
REFERENCES
\end{center}
$[^1]$ {\sc Howard, L. N.} J. Fluid Mech.  {\bf 17}, 1963, No. 1, 405--432.\\
$[^2]$ {\sc Busse, F. H.} J. Fluid. Mech.  {\bf 37}, 1969, No. 1, 457--477.\\
$[^3]$ {\sc Doering C. R., P. Constantin.} Phys. Rev. Lett. {\bf 69}, 1992, No. 11,  1648 -- 1651.\\ 
$[^4]$ {\sc Vitanov, N. K., F. H. Busse.} Phys. Rev. E {\bf 63}, 2001, Issue 1, Part 2, 
	Article Number 016303.\\
$[^5]$ {\sc Hoffmann, N. P, N. K. Vitanov.} Phys. Lett. A  {\bf 255}, 1999, Nos. 4-6, 277--286.\\
$[^6]$ {\sc Strauss J. M.} Physics of Fluids {\bf 17}, 1974, No. 3, 520 -- 527.\\
$[^7]$ {\sc Vitanov, N. K.} Physica D  {\bf 236}, 2000, Nos. 3--4, 322--339.\\
$[^8]$ {\sc Vitanov, N. K., F.H. Busse}, Zeitschrift fur Angewandte Mathematik
	und Physik (ZAMP) {\bf 48}, 1997, No. 2, 310--324.\\
$[^9]$ {\sc Vitanov, N. K.} Phys. Lett. A  {\bf 248}, 1998, Nos. 5--6, 338--346.\\
$[^{10}]$ {\sc Vitanov, N. K.}  Physics of Fluids {\bf 17}, 2005, No. 10, Article Number 105106.\\
$[^{11}]$ {\sc Vitanov, N. K.} European Physical Journal B {\bf 23}, 2001, No. 2, 249--266.\\
$[^{12}]$ {\sc Vitanov, N. K.} Phys. Rev. E {\bf 62}, 2000, No. 3, 3581--3591.\\
$[^{13}]$ {\sc Vitanov, N. K.} European Physical Journal B {\bf 15}, 2000, No. 2, 349--355.\\
$[^{14}]$ {\sc Vitanov, N. K.} Phys. Rev. E  {\bf 67}, 2003,  Issue 2, Part 2, Article Number 026322.\\
$[^{15}]$ {\sc Vitanov, N.} Compt. rend. Acad. bulg. Sci., {\bf 63}, No. 5, 2010, 685--692.\\
$[^{16}]$ {\sc Radev S., N. Vitanov.} Compt. rend. Acad. bulg. Sci., {\bf 64}, No. 3, 2011, 353 -- 360.\\
$[^{17}]$ {\sc Vitanov, N. K.} Phys. Rev. E {\bf 61}, 2000, No. 1, 956--959.\\
$[^{18}]$ {\sc Vitanov, N. K.} European Physical Journal B {\bf 73}, 2010, No. 2, 265--273.\\
$[^{19}]$ {\sc Chandrasekhar, S.} Hydrodynamics and Hydromagnetic Stability. Dover, New York, 1981.\\

\begin{flushleft}
{\sl $^1$ 'G. Nadjakov' Institute of Solid State Physics\\
Bulgarian Academy of Sciences \\
Blvd. Tzarigradsko Chaussee 72, 1784 Sofia, Bulgaria \\
e-mail: zdim@issp.bas.bg}
\end{flushleft}
\begin{flushleft}
{\sl $^2$ Institute of Mechanics, Bulgarian Academy of Sciences \\
Acad. G. Bonchev Str., Bl. 4, 1113 Sofia, Bulgaria \\
e-mail: vitanov@imbm.bas.bg}
\end{flushleft}

\end{document}